\documentclass[floatfix,twocolumn,showpacs,preprintnumbers,amsmath,amssymb,prb]{revtex4}
\usepackage{latexsym}
\usepackage{graphicx}
\usepackage{dcolumn}
\usepackage{bm}

\begin{document}

\title{Band structure calculations for Ba$_{6}$Ge$_{25}$ and Ba$_{4}$Na$_{2}$Ge$_{25}$ clathrates}
\author{Ivica Zerec}
\author{Alexander Yaresko}

\author{Peter Thalmeier}
\author{Yuri Grin}
\affiliation{
Max Planck Institute for Chemical Physics of Solids, D-01187 Dresden,
Germany
}
\date{\today}

\begin{abstract}

Electronic band structures for Ba$_{6}$Ge$_{25}$ and
Ba$_{4}$Na$_{2}$Ge$_{25}$ clathrates are calculated using linear
muffin-tin orbital method within the local density approximation. It
is found that barium states strongly contribute to the density of
states at the Fermi level and thus can influence the transport properties of
the compounds. A sharp peak of the density of states is
found just at the Fermi level. It is also shown that the shifting of barium atoms toward
experimentally deduced split positions in Ba$_{6}$Ge$_{25}$ 
produces a splitting of this peak which may be interpreted as a band
Jahn-Teller effect. If the locking of the barium atoms at the observed
structural phase transition is assumed, this reduction of the density 
of states at the Fermi level can account for the experimentally
observed decrease of the magnetic susceptibility and electrical
resistivity at the phase transition, and the values of density of
states are in agreement with low temperature specific heat
measurements and variation of superconducting transition temperature
with pressure.

\end{abstract} 


\maketitle

\section{Introduction}

Ba$_{6}$Ge$_{25}$ clathrate\cite{Ba6Ge25,Kim,Fukuoka} attracted considerable 
scientific interest and initiated an intense experimental activity because 
of its rich physical properties. This class of compounds is
investigated primarily because of their thermoelectric properties.\cite{Nolas} The 
concept of PGEC, phonon-glass electron-crystal, was introduced. Massive 
alkali and alkali-earth metal atoms are expected to rattle in oversized
cages usually formed by Si, Ge or Sn atoms. Thus, they scatter the lattice
phonons and reduce the thermal conductivity. On the other hand,
electrical conductivity is believed to be more constrained to the cage 
and not much affected by rattling.

Ba$_{6}$Ge$_{25}$ undergoes a two-step first order
structural phase transition at temperatures 215 and 180 K.
\cite{Paschen,Silke}
The structural phase transition is accompanied by dramatic change of
transport properties. Above the transition it shows a
metal-like conductivity with fairly large value for this class of
compounds. At the phase transition this behaviour is profoundly
changed; resistivity abruptly increases, then slowly rises with
lowering the temperature and saturates at low
temperatures. The Hall coefficient smoothly decreases with decreasing the
temperature and is not remarkably affected by the phase
transition. Magnetic susceptibility is negative and temperature independent
above and abruptly decreases at the phase transition similarly to the
thermopower. \cite{Paschen,Silke} It has been observed that hydrostatic
pressure suppresses the phase transition.\cite{Malte} The homologue
clathrate Ba$_{4}$Na$_{2}$Ge$_{25}$,\cite{Na2Ba4Ge25} with smaller
lattice parameter, does not undergo a phase transition and shows typical
metal-like behaviour.\cite{Silke} At low temperatures both
clathrates experience superconducting transitions. \cite{Malte}

In this article we consider the  electronic band structure of the compounds and the 
influence of the structural phase
transition on the electronic band structure as well as the possible
consequences on the transport properties. In Section II some
structural details are considered. Details of the calculations are
described in Section III. The results and the discussion are presented in
Section IV. Finally, the summary and the conclusions are given in Section V.

\section{Structure}

The structure of Ba$_{6}$Ge$_{25}$ belongs to the space group $P4_{1}32$ (No.~213). There
are 124 atoms in the unit cell. All the structural details can be
found in Ref.~\onlinecite{Ba6Ge25,Na2Ba4Ge25,Grin,Grin2}. Thorough description of the structure can also be
found in Ref.~\onlinecite{Fassler2,Fassler} on structurally similar Sn clathrates. Germanium atoms form the framework which defines
three different types of cages filled with barium atoms. The main
building blocks are distorted germanium pentagon-dodecahedra with Ba1
atoms inside. They form a chiral network with large cavities,
where Ba2 and Ba3 atoms are placed. In the Ba$_{4}$Na$_{2}$Ge$_{25}$, two sodium atoms are randomly distributed in
three Ba2 sites per formula unit. In total, 17 of 25 germanium atoms
per formula unit are four-fold coordinated by other germanium atoms. The
remaining 8, usually denoted as Ge3 and Ge5, are three-fold
coordinated. The average distance between germanium atoms ($\sim$2.54~\r{A}) is somewhat
larger than for Ge in diamond type structure ($\sim$2.45~\r{A}) and the angles 
are distorted. Typical nearest distance between
barium and germanium atoms is around 3.5~\r{A}. The nearest
distance between barium atoms is around 4.70~\r{A} for Ba2-Ba3 in
undistorted structure (see also Ref.~\onlinecite{Grin2}). 

Due to highly aspherical electron density deduced from X-ray
diffraction data, split positions for Ba2 and Ba3 atoms are
assumed. The most pronounced effect at the
structural phase transition is the increase of the atomic displacement
parameter for Ba2, and somewhat less for Ba3 atom.
\cite{Grin,Grin2} It is believed that below the phase transition Ba2
and Ba3 atoms are 
actually locked in the split sites. However, since no superstructure
is observed, they should be randomly distributed among split sites. The germanium cage also undergoes modifications at the phase
transition and there is not only a difference between the room temperature and
the low temperature structure, but also the variation with temperature
through the whole temperature range.\cite{Grin2}

\begin{table*}
\caption{\label{symm} Wyckoff positions splitting from the group No.~213 
  in its subgroups.\cite{Muller} The average structure of
  Ba$_{6}$Ge$_{25}$ belongs to 
  the space group $P4_{1}32$ (No.~213). The values for the atomic
  coordinates were taken from Ref.~\onlinecite{Ba6Ge25,Grin2}. Shifting the Ba2 atom toward
  one of the split positions of Ba4 site demands lowering the symmetry to the space group
  $P2_{1}3$ (No.~198). For Ba$_{4}$Na$_{2}$Ge$_{25}$ (atomic
  coordinates were taken from Ref.~\onlinecite{Na2Ba4Ge25}) the symmetry had to be 
  lowered to the group $P4_{1}2_{1}2$ (No.~92) so that Na atoms could
  be orderly placed on Ba2 sites. Shifting the Ba or Na from Ba2 site
  demands lowering the symmetry to the space group $P4_{1}$ (No.~76).} 
\begin{tabular}{|c||c|c||c|c||c|c||c|c|}\hline
Sites & \multicolumn{2}{c||}{No.~213} & \multicolumn{2}{c||}{No.~198} &
 \multicolumn{2}{c||}{ No.~92} & \multicolumn{2}{c|} {No.~76} \\
 \hline\hline
Ge1  &  & x, y, z & & x, y, z & & x, y, z & 4\textit a & x, y, z \\ \cline{8-9}
Ge5 &  & & 12\textit b & &\raisebox{1.5ex}[-1.5ex]{8\textit b} & & 4\textit a & -x+1/2, y+1/2,
  -z+1/4 \\ \cline{6-9}
Ge6 & & & & & & z+7/8,x+3/4,y+3/8 & 4\textit a & z+7/8,x+3/4,y+3/8 \\ \cline{4-5}
 \cline{8-9}
Ba4 (Na)\footnote{Split positions of the Ba2 and Na.} &\raisebox{1.5ex}[-1.5ex]{24\textit e} & & & y+3/4, x+1/4, -z+1/4 &
\raisebox{1.5ex}[-1.5ex]{8\textit b} & & 4\textit a & -z-3/8, x+1/4, -y-1/8 \\
\cline{6-9}
& & & 12\textit b & & & y+1/4, z+5/8, x+1/8 & 4\textit a & y+1/4, z+5/8, x+1/8 \\
 \cline{8-9}
& & & & & \raisebox{1.5ex}[-1.5ex]{8\textit b} & & 4\textit a & -y+1/4, z+1/8, -x+1/8
 \\ \hline
Ba2 (Na)& & 1/8, y, y+1/4 & & 1/8, y, y+1/4 & & 7/8, y+1/2, y+1/8 & 4\textit a & 7/8,
 y+1/2, y+1/8 \\ \cline{8-9}
Ge4 & 12\textit d & & 12\textit b & & \raisebox{1.5ex}[-1.5ex]{8\textit b} & & 4\textit a & -3/8, y, -y+1/8
\\ \cline{6-9}
& & & & & 4\textit a & y+3/4, y+3/4, 0 & 4\textit a & y+3/4, y+3/4, 0 \\ \hline
Ba1& & x, x, x & 4\textit a & x, x, x & & x+3/4, x+1/2, x+7/8 & 4\textit a & x+3/4, x+1/2,
 x+7/8 \\ \cline{4-5} \cline{8-9}
Ge2, Ge3 & \raisebox{1.5ex}[-1.5ex]{8\textit b} & & 4\textit a & -x+3/4, -x+3/4, -x+3/4 &
\raisebox{1.5ex}[-1.5ex]{8\textit b} & & 4\textit a & -x-1/4, x, -x-5/8 \\ \hline
Ba3 & 4\textit a & 3/8, 3/8, 3/8 & 4\textit a & 3/8, 3/8, 3/8 & 4\textit a & 1/8, 7/8, 1/4 & 4\textit a & 1/8,
7/8, 1/4 \\ \hline
 
\end{tabular}
\end{table*}

\section{Computational details}

Band structure calculations were performed within Local Density
Approximation (LDA) using Linear Muffin Tin
Orbital (LMTO) method in Atomic Sphere Approximation (ASA) with
combined correction terms taken  into account.\cite{Andersen, Skriver} Scalar
relativistic version was used with von Barth-Hedin
exchange-correlation potential.\cite{Barth}

The calculations for the room- and the low-temperature structural data for 
Ba$_{6}$Ge$_{25}$ were performed. The
atomic parameters for different models used in calculations are given in
Table~\ref{symm}. The barium atoms were placed in their 
average positions within higher symmetry model, i.e., space group
$P4_{1}32$. In Ba$_{4}$Na$_{2}$Ge$_{25}$, the random distribution
of the Na and Ba atoms at Ba2 site was neglected and the calculations
were performed for lower symmetry model, namely,
group $P4_{1}2_{1}2$ (No.~92), with sodium atoms at one of the 8$b$ symmetry
sites.

The calculations with Ba2 and Na atoms shifted from their average positions in the
original space groups (No.~213 and No.~92, respectively) were performed using lower symmetry
models in space groups $P2_{1}3$ (No.~198) and $P4_{1}$ (No.~76), respectively. In
this way the assumed randomness of the locking was not considered.

Most of the results presented in this work were obtained from
self-consistent calculations on $12\times12\times12$ k-space mesh. For
the detailed band structure around Fermi level the calculations were
performed for the $16\times16\times16$ mesh. The values of the DOS at
the Fermi level for  these cases showed corresponding variations of
about 1~$eV^{-1}$ per unit cell. The total energy was
negligibly affected by this change of the mesh size.

The atomic sphere radii were determined so that their overlap was
minimised. This resulted in the radii of about 1.5~\r{A} (2.9~a.u.) for germanium and
2.5~\r{A} (4.8~a.u.) for barium atoms. It is worth to note that, as the radii of germanium
spheres are rather small, germanium $p$-states extend far beyond the sphere
boundaries. The large part of these states, so called "tails," reside
in the neighbouring barium spheres. This should be taken into account when
considering barium and germanium charges which are defined as
the integrals of the electron density over the corresponding sphere.
They are clearly identified in the empty spheres of hypothetical
$\Box_{6}$Ge$_{25}$, where the barium atoms are replaced by the empty
spheres of the same radii. In this case a high electron density near the
boundary can be observed which diminishes toward the center of the
empty sphere. Because of the large barium spheres, also the states of
$f$-symmetry were used in the expansion of the wave function. In the germanium spheres the wave
function was expanded up to the states of $d$-symmetry.

\section{Results and discussion}

Here we present and discuss the results for the low temperature
structure of Ba$_{6}$Ge$_{25}$ at 10~K, first with Ba atoms in the average 
sites. We focused the analysis on the density of states and the bonding
properties between the barium atoms and the germanium cage. No
distinctive qualitative difference in the DOS between the room  
and the low temperature structures within higher symmetry model was found.
Further, we investigated the effects of shifting the 
barium atoms toward split positions on the band structure. This was done within
the low temperature structure, because there the split positions are
the most distant and the locking of the barium atoms is assumed. Finally,
the results for the Ba$_{4}$Na$_{2}$Ge$_{25}$, which does not undergo
phase transition, are discussed. In particular, the influence of the
sodium on the band structure is examined in comparison with barium.

\subsection{Electronic band structure and density of states}

\begin{figure}
\includegraphics[width=.47\textwidth,clip=]{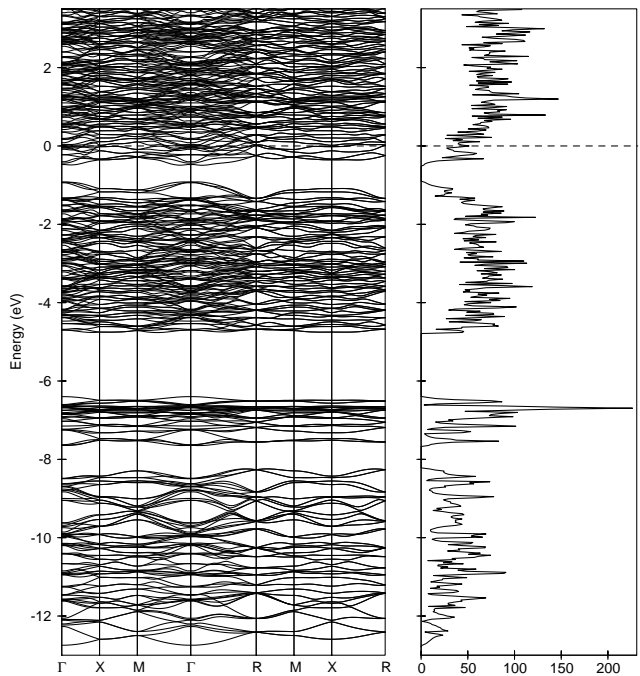}
\caption{\label{lt12}The electronic band structure and the density of 
  states for
  Ba$_{6}$Ge$_{25}$ in the low temperature structure at 10~K with Ba
  atoms in the average positions within space group $P4_{1}32$.}
\end{figure}

The electronic band structure with its density of states
calculated for Ba$_{6}$Ge$_{25}$ in the low temperature undistorted
structure are shown in
Fig.~\ref{lt12}.
The overall picture is similar for the room temperature structure.
Due to the large number of atoms in the unit cell there is a large number
of bands and a complex electronic band structure. 

\begin{figure}
\includegraphics[width=.45\textwidth,clip=]{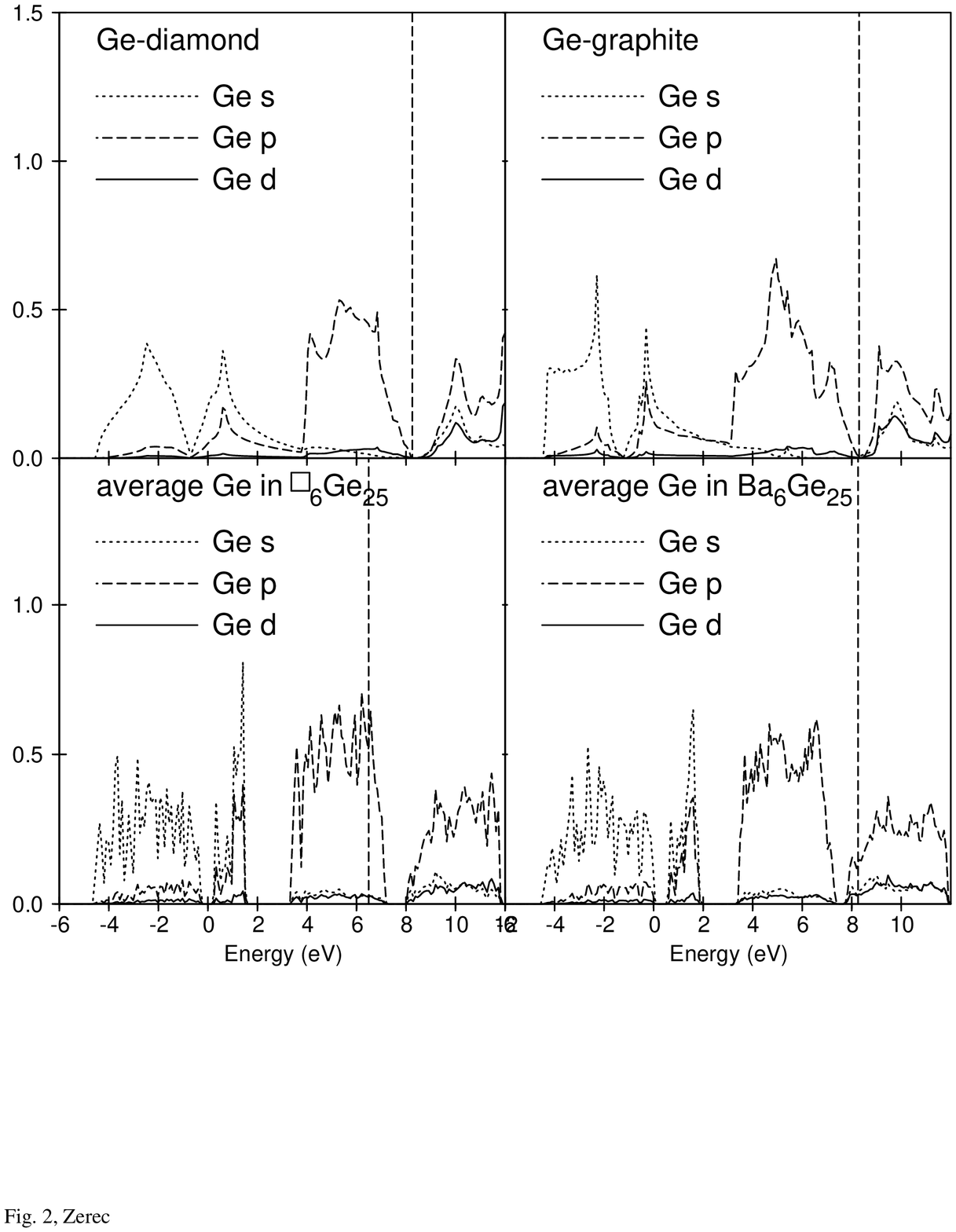}
\caption{\label{gegege}Projected density of states from Ge atomic
  spheres in Ge-diamond, hypothetical graphite structures,
  Ba$_{6}$Ge$_{25}$ and $\Box_{6}$Ge$_{25}$. For the clathrate
  compounds, the average contribution from all germanium is shown.}
\end{figure}

In Fig.~\ref{gegege}
the $l$-projections of DOS in germanium atomic spheres are shown: Ge in
diamond structure, hypothetical Ge-graphite, $\Box_{6}$Ge$_{25}$,
and Ba$_{6}$Ge$_{25}$ compounds are compared. In the DOS, $s$-, $sp$-, and $p$-bonding regions can
be recognised in filled and empty clathrate similar to those in Ge-diamond or
graphite. This similarity is to be expected since locally 17 of 25
germanium atoms per formula unit are four-fold coordinated (4b) and form
covalent $sp^3$-like bonds, like in the diamond structure. The other 8 
germanium atoms are three-fold bonded (3b) more like in graphite structure.
The total width of the bonding region is about 0.9 eV narrower in
the empty and filled clathrates. The fundamental gap
between bonding and anti-bonding region is larger in
$\Box_{6}$Ge$_{25}$ than in the diamond structure by the same amount
of 0.9 eV and there appears a gap between $sp$- and $p$-like bands in the valence
region. These features are common for the other Ge and Si clathrates
(e.g., Ref.~\onlinecite{Dong}). The appearance of the gap in the
valence region by its contraction was associated with
the 5-ring patterns (pentagons) of germanium or silicon atoms,\cite{Saito,Moriguchi}
but there was also some criticism of the
idea.\cite{Moriguchi2} According to the latter, the small angular
distortion of the tetrahedrally bonded framework
 may also play an important role. While for Ge clathrates of type I
 and II an overlap of the $s$- and $sp$-region is observed,\cite{Dong} here
one can see a clear separation. One should
keep in mind that in Ba$_{6}$Ge$_{25}$ structure, in contrast to the
clathrates I and II, there are three bonded germanium atoms. However, the
band structure calculation on hypothetical Ge-graphite, where all germanium
atoms are three-fold coordinated, did not reveal pronounced tendency
toward such a separating of the $s$- and $sp$-regions, as seen in
Fig.~\ref{gegege}.

The main difference in the site projected DOS for germanium atomic spheres between the
empty and the filled clathrate is the reduction of the gap
between bonding and anti-bonding region in the filled clathrate. This
immediately shows the importance of barium atoms not only as the pure
electron donors to the conduction band formed of germanium anti-bonding states.

The three-bonded germanium atoms, six Ge5 and two Ge3, form a distorted cube
around Ba3. Ba2 atoms lie on the lines passing from
the central Ba3 through the cube faces formed by three Ge5 and one Ge3
atom. In this way they form a large octahedron around Ba3. The closest 
distances are between Ba2 and Ge3, 3.35~\r{A}, and Ba3 and Ge5
3.38~\r{A}. Ge3 atoms are somewhat more distant from Ba3, at 
3.89~\r{A}. This environment and the nature of the bonds between Ba3,
Ge3, Ge5, and Ba2 should be the key to understand the physical
properties of the compound and its phase transition.\cite{Grin2}

\begin{figure}
\includegraphics[width=.45\textwidth,clip=]{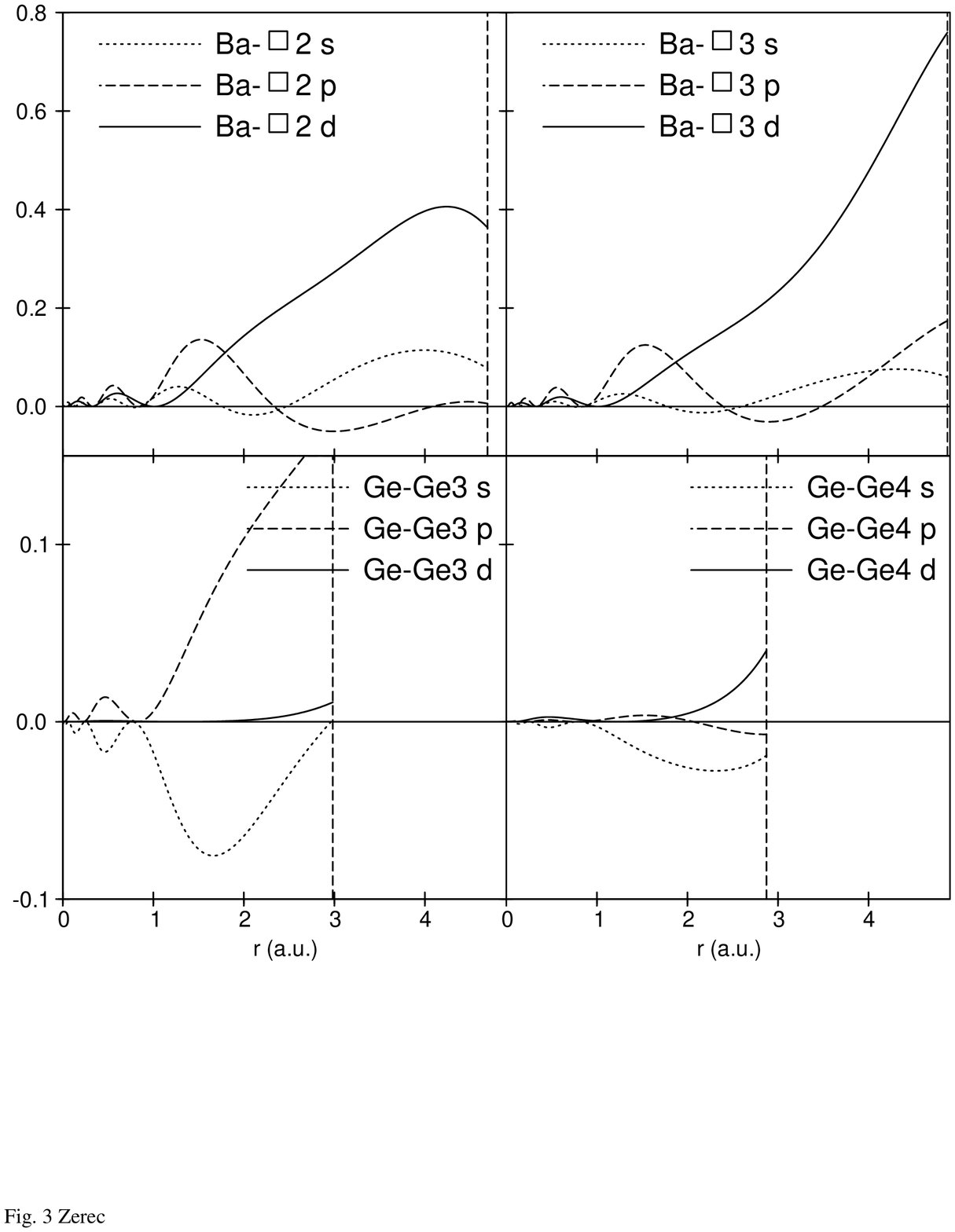}
\caption{\label{diffden}Electron density difference between filled and 
  empty clathrate in particular atomic spheres.}
\end{figure}

Let us consider the charges in the atomic spheres. As
mentioned in Section III, there is a large contribution of germanium states
in the oversized barium spheres. This can be seen in Table~\ref{charge} in
the column for the charges of empty clathrate, $\Box_{6}$Ge$_{25}$.
\begin{table}
\caption{\label{charge}Total charges in the atomic spheres and their radii 
  for Ba$_{6}$Ge$_{25}$ and $\Box_{6}$Ge$_{25}$.}
\begin{ruledtabular}
\begin{tabular}{lcccc}
At. Sphere & Ba$_{6}$Ge$_{25}$ & $\Box_{6}$Ge$_{25}$ & difference & AS
radius (a.u.)\\
\hline
Ba1, $\Box1$ & 4.613 & 2.853 & 1.760 & 4.7175 \\
Ba2, $\Box2$ & 3.474 & 1.864 & 1.610 & 4.6919 \\
Ba3, $\Box3$ & 3.711 & 1.776 & 1.935 & 4.8720 \\
Ge1 & 3.520 & 3.462 & 0.058 & 2.8663 \\
Ge2 & 3.821 & 3.805 & 0.016 & 2.9607 \\
Ge3(3b) & 3.536 & 3.403 & 0.133 & 2.9829 \\
Ge4 & 3.613 & 3.619 & -0.006 & 2.8728 \\
Ge5(3b) & 3.436 & 3.315 & 0.121 & 2.8675 \\
Ge6 & 3.559 & 3.499 & 0.060 & 2.8648 \\
\end{tabular}
\end{ruledtabular}
\end{table}
The large charges in the empty spheres are coming from the tails of germanium
states. The highest value is in $\Box$1. This is expected because
$\Box$1 has the closest Ge neighbours forming distorted
pentagon-dodecahedron. Despite of the fact that $\Box$3 has the largest 
sphere, relatively smaller charge is found inside due to the fewer Ge
neighbours. For three-bonded germanium atoms it can also be noted that they
have somewhat smaller charges inside their spheres in comparison to
the four-bonded germanium, although e.g., Ge3 has the largest atomic
sphere from all germanium atoms. Among germanium atoms, three-bonded Ge3 and Ge5
have the largest increase of the charge upon inclusion of barium atoms. This is in
accord with the expectation of Zintl concept, namely, that three-bonded germanium
atoms accept an electron from barium atoms to form the lone electron
pairs. However, because of the small germanium spheres the absolute value of
the difference is not so pronounced. Among barium spheres, the largest
increase of the charge, in comparison to the corresponding empty spheres,
undergoes the Ba3 sphere. We interpret this as the influence of
the lone pairs, reflected in the Ba3 sphere as the tails from
neighbouring three-bonded germanium atoms. This is illustrated in
Fig.~\ref{diffden} where the differences of the $l$-projected radial
charge densities in the several atomic spheres are
shown. The large electron density near the boundary of Ba3 atomic
sphere may be associated with the lone electron pairs of the neighbouring
three-bonded germanium atoms. It was also shown  by
electron localisation function (ELF) analysis that these lone pairs 
are situated around Ba3 site.\cite{Grin2} The oscillating part of the 
density, i.e., its difference shown in Fig.~\ref{diffden}, may be interpreted as more atomic-like barium states. The biggest
contribution comes from the states of $d$-symmetry.

We conclude that there is only a small charge transfer from barium to the cage,
just in agreement with conclusion made for
Ba$_{8}$Ga$_{16}$Ge$_{30}$ in Ref.~\onlinecite{Blake2}, with
difference, that here all the arguments point out that a considerable
amount of the electron charge remains in the states with significant
barium character. This is in contrast  to the situation reported for
K$_{8}$Ge$_{46}$ in Ref.~\onlinecite{Zhao} where the
authors conclude complete charge transfer from potassium atom to the
frame. Both of these cited results were critically re-considered and charge
transfer redefined in the recent paper. \cite{Nagano} However, as may be seen
in Fig.~\ref{diffden}, the large amount of the charge in the barium
spheres resides near the boundaries, so that it may be said that the conduction
electrons are more constrained to the cage.

\begin{figure}
\includegraphics[width=0.45\textwidth,clip=]{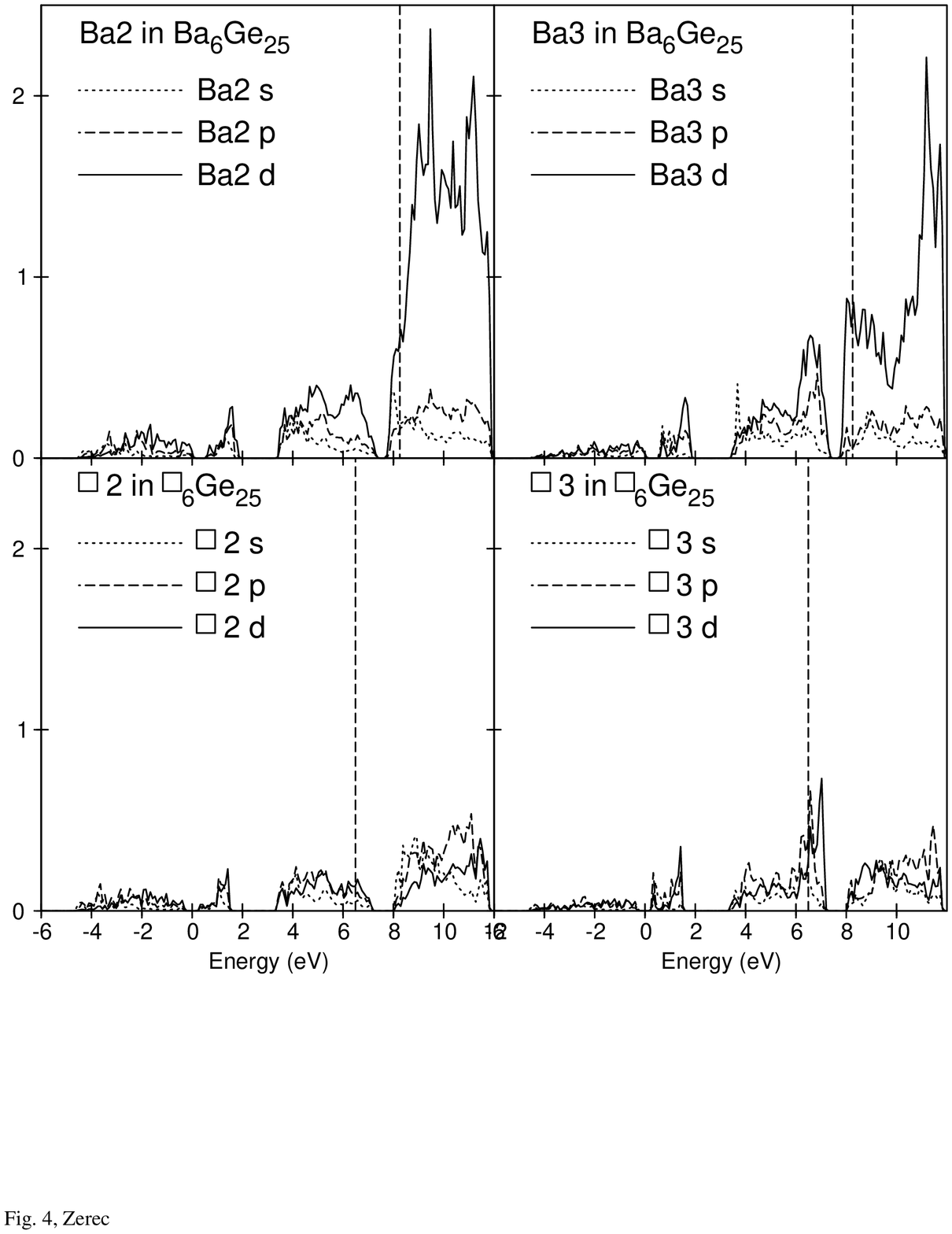}
\caption{\label{eba}The projected density of states from atomic
  spheres of Ba2 and Ba3 in Ba$_{6}$Ge$_{25}$ and the equivalent empty
  spheres of $\Box$2 and $\Box$3 in $\Box_{6}$Ge$_{25}$.} 
\end{figure}

In the band structure calculations for the similar K$_{6}$Sn$_{25}$ and
K$_{6}$Sn$_{23}$Bi$_{2}$ compounds, a lifting of the three
bands, originated only from three-bonded tin atoms, from the top of the
valence region was found and ascribed to the interaction of the lone pairs
formed by three-bonded tin atoms. \cite{Fassler}. But, as was pointed out there, in
Ba$_{6}$Ge$_{21}$In$_{4}$ this characteristic is absent due to a smaller
spatial extension of germanium orbitals. We did
not notice significant difference of DOS between three-bonded (Ge3, Ge5) and
four-bonded germanium atomic spheres. However, as discussed above, we expect that 
the lone electron pair states mostly reside as
tails in $\Box$3 atomic sphere. Therefore one might expect their
contribution to the DOS in $\Box$3 sphere in
$\Box_{6}$Ge$_{25}$. Indeed, we find a peak in DOS on the top of the
unoccupied valence region (Fig.~\ref{eba}) appearing only in $\Box$3
sphere. This peak remains in the filled clathrate, i.e., in Ba3
sphere, what can also be seen in Fig.~\ref{eba}. For Ba2 and Ba3 spheres a
large increase of DOS, compared to the empty $\Box$2 and $\Box$3 
spheres, in the region around the Fermi energy is found. The number
of states is increased and this suggests that the barium states participate in 
the conduction band formation.

This is in contrast to the results for the Ba$_{8}$Ga$_{16}$Ge$_{30}$
clathrate \cite{Blake2} where the inclusion of
barium atoms inside the Ga-Ge cage left the band structure
almost intact. This is also different to the cases of some Si
clathrates \cite{Demkov} where the main
effect of the inclusion of alkali metal was shifting of the particular
state at the bottom of the conduction band, but still predominantly of
the cage-like character so that it was possible to neglect the overlap
of the orbitals of guest atoms with host-cage. Here, however, the
distances between the barium and germanium atoms are shorter and stronger
hybridisation may be expected.

\begin{figure}
\includegraphics[width=.35\textwidth,clip=]{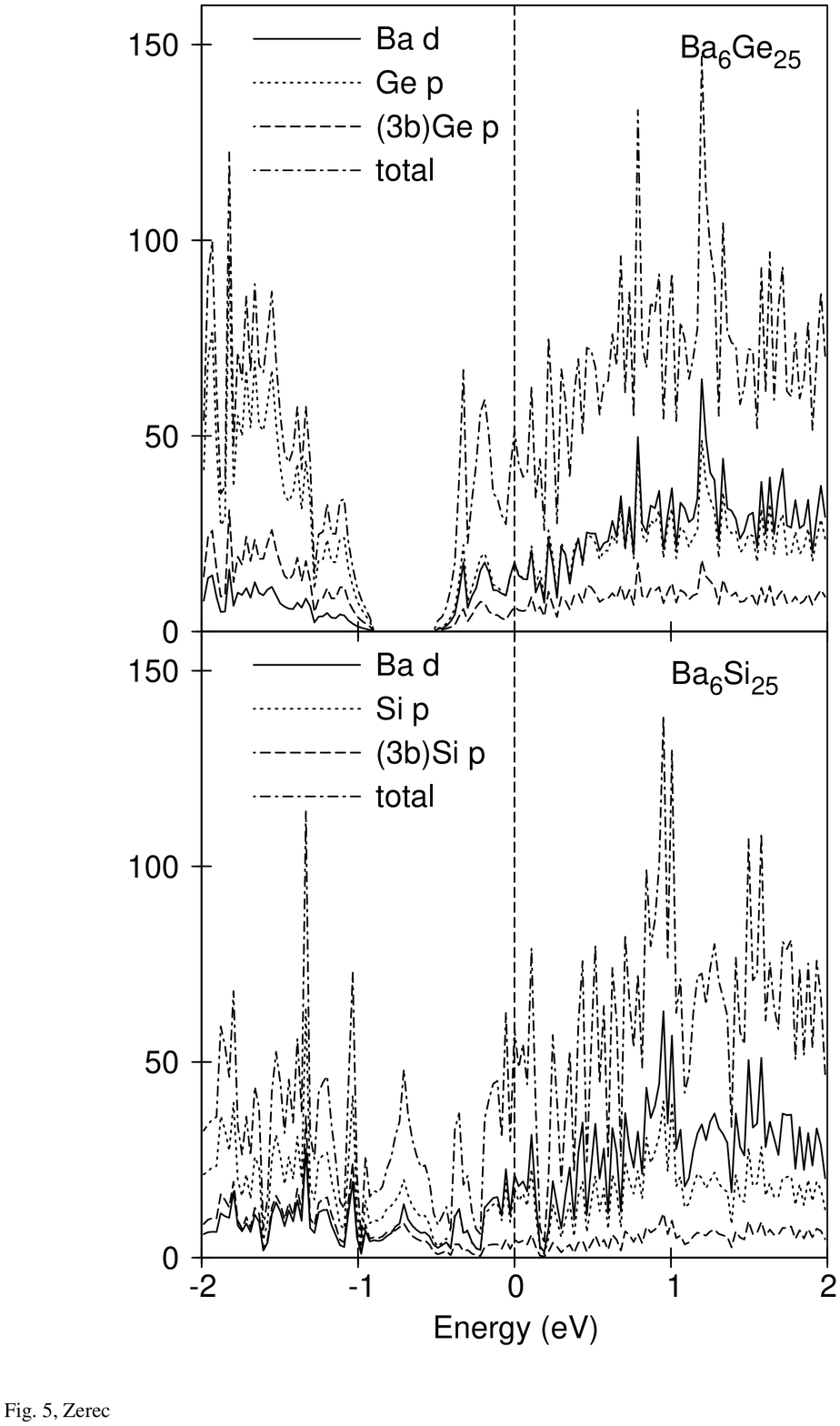}
\caption{\label{dostot}Contribution to DOS from all Ba $d$-states,
  all Ge $p$-states and all 
  three-bonded Ge $p$-states per formula unit. The total DOS is also
  shown. In the figure below the same is shown for the case of
  Ba$_{6}$Si$_{25}$, with Si atoms instead of Ge.}
\end{figure}

Another important feature of DOS is that it shows a
peak just at the Fermi level (Fig.~\ref{dostot}). Its value is around 50 states/eV per
unit cell in the low temperature
structure. The largest contributions come from barium atomic
spheres mainly with $d$-symmetry and from $p$-states in germanium atomic spheres. The shape of the DOS coming from 
barium and germanium spheres follows closely each other. This indicates a strong
hybridisation of barium and germanium states. The contribution from Ba1
spheres is significantly lower than that from Ba2 and Ba3 spheres, although
Ba1 has the nearest Ge neighbours. This
complies with the assumption that the Ba2 and Ba3 atoms with the
surrounding three-bonded germanium atoms determine the most unusual properties of
the system. It is quite interesting that the similar peak of the DOS at
the Fermi energy was found for Ba$_{8}$Si$_{46}$
clathrate \cite{Moriguchi,Saito} as well. In that case also the
relatively shorter distances between the barium and silicon atoms were found
($\sim$3.3-3.5~\r{A}). We have also performed calculation for
Ba$_{6}$Si$_{25}$.\cite{Fukuoka2} The lattice parameter is smaller
compared to the Ge clathrate and the fundamental gap disappears
completely but the peak in the DOS at the Fermi level with strong
contribution of the states with $d$-symmetry coming from the barium atomic
spheres remains (Fig.~\ref{dostot}). It is also interesting that the
subband-gap between $s$- and $sp$- regions (not shown in the
Fig.~\ref{dostot}) appears in the Ba$_{6}$Si$_{25}$ as well.

For a better understanding of the above results, i.e., the role that is played
by barium states, we have performed the 
calculation with barium atoms only, Ba$_{6}\Box_{25}$, without the Ge
framework. The
distances between Ba2 and Ba3 atoms of about 4.70~\r{A} are comparable
to the distances between atoms in Ba bcc structure with 4.34~\r{A}. The
calculation showed the dominant contribution of $d$-symmetry states at
the Fermi level. The $s$-states are split into bonding and the anti-bonding
regions. The bonding states lie below Fermi level and
anti-bonding above, so that they do not give significant contribution to 
the DOS at the Fermi level. The tails of the barium $d$-states in
the empty germanium
spheres are of $s$- and $p$-symmetry, so that the appreciable overlap of these
states with germanium $s$- and $p$-states in the filled clathrate might be
expected. Concerning the relative 
energy parameters of the barium and germanium muffin-tin orbitals, they showed
qualitative agreement with the values used for the atomic orbitals in LCAO
calculations. \cite{Harrison} All these facts suggest that the appearance of
the barium $d$-states and their hybridisation with germanium
$p$-states at the Fermi level is quite plausible.

The possible importance of the hybridisation of the Ba 5$d$ states with
Si cage was also pointed out for  Ba$_{8}$Si$_{46}$.\cite{Saito} The results of
the experiment with ultrahigh-resolution photoemission spectroscopy
for Ba$_{8}$Si$_{46}$ suggest even stronger hybridisation of Ba and Si
states than obtained in LDA calculation. \cite{Yokoya} NMR experiments
on this compound could be interpreted if strong contribution of Ba
$d$-states to the DOS at the Fermi energy is assumed. \cite{Shimizu}
 
\subsection{Band structure near the Fermi level versus Ba2 shift}

The presence of the peak in the DOS at the Fermi level suggests a possible
instability of the system. Since there are strong indications of
locking of Ba2 atoms at the phase transition, as mentioned in Section
II, we investigated the modifications of the band structure due to the
shifting of Ba2 atoms within the low temperature structure.

\begin{figure}
\includegraphics[angle=-90,width=.45\textwidth,clip=]{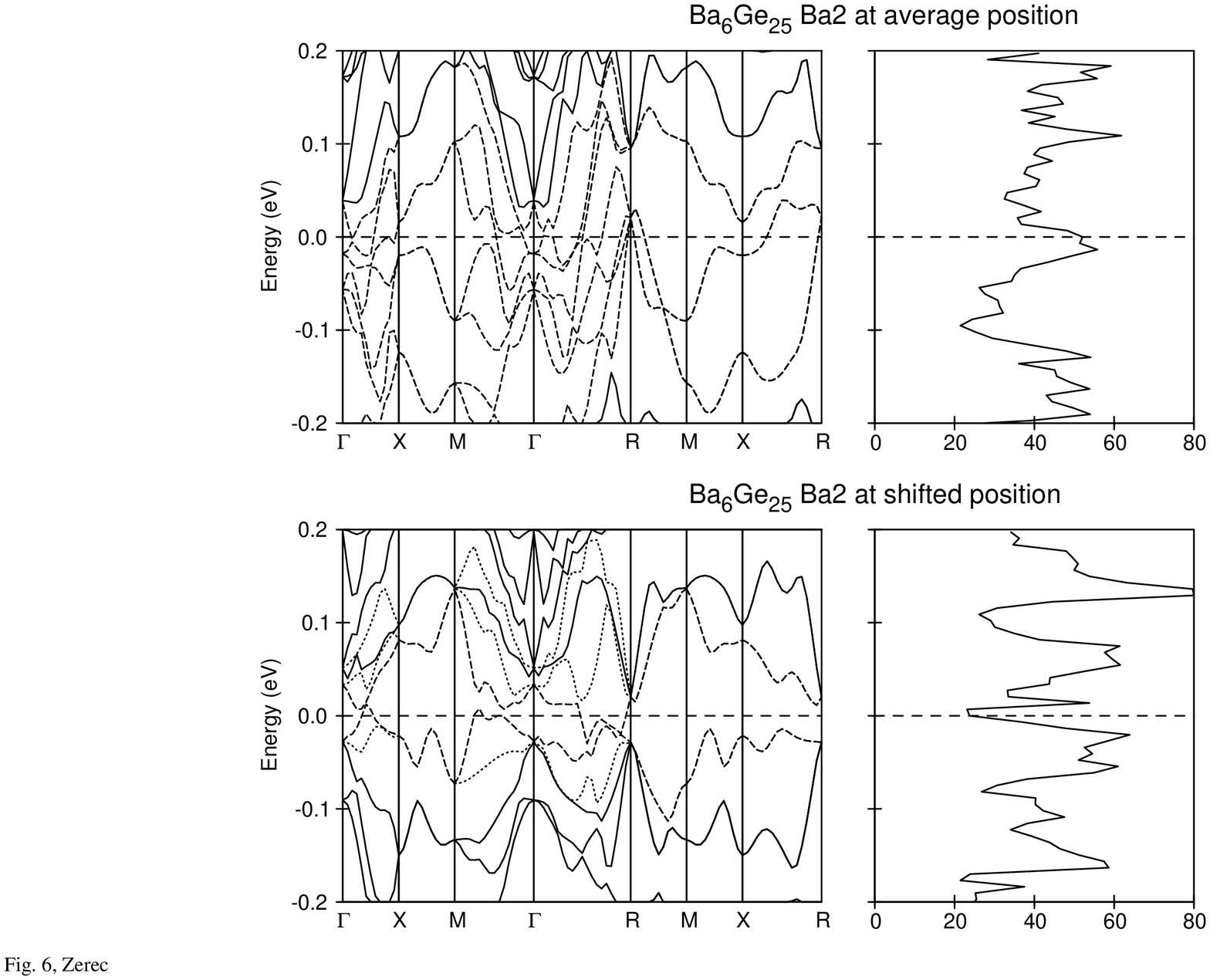}\\
\caption{\label{zoom} Electronic band structure and the density of
  states  around the
  Fermi energy for the Ba2 at the average position and at 80\%
  toward the split site in
  Ba$_{6}$Ge$_{25}$. Dashed lines designate the bands that are
  crossing the Fermi level.}
\end{figure}

The most interesting effect of this shifting of Ba2 atoms, performed as
explained in Section III, is the splitting of the peak at the
Fermi level and, consequently, the reduction of DOS. The effect was
found to be the most pronounced with Ba2 at 80\% distance from the
higher symmetry  position toward the experimental split position. In
Fig.~\ref{zoom}, the band structure together with the DOS in the close
vicinity of the Fermi level for this structure is compared with the
higher-symmetry (average) structure. The effect may be considered as a band
Jahn-Teller effect, though in the band structure along the symmetry
directions no obvious removing of the degeneracy is observed. The
equivalent scenario was proposed for some other compounds with
somewhat similar metal-insulator transition. \cite{Hagino} The
conventional Jahn-Teller effect was considered also for some other
clathrates. \cite{Demkov,Tse,Brunet}

The total energy was also compared for shifted positions. The minimum
was found to be at the off average position. The potential
energy surface for Ba atoms showed binding stronger than expected, but
the variation of the total energy with respect to the shifting of
atoms is known not to be highly accurate within the atomic sphere
approximation. That is also the reason why the structure optimization
was not performed. 

One has to be careful with the interpretations, especially because the
whole framework is slightly changing with the temperature and
is strongly disturbed during the phase transition. \cite{Grin2} As
mentioned before, we did not notice a
qualitative difference in the DOS between the room temperature and the low
temperature higher symmetry 
structures. For both of them the peak at the Fermi level is
found. The splitting of the peak when Ba2 atom is shifted toward one of
the split positions indicates that the locking of barium atoms might
be the most important
process at the phase transition. If we assume that at the phase
transition this locking of barium atoms occurs then the
reduction of the DOS at the Fermi level is in a qualitative agreement
with the observed decrease of magnetic susceptibility. The decrease of 
the value of DOS is found to be around 25~$eV^{-1}$ per unit cell which 
corresponds to a drop of the Pauli susceptibility of
$2.08\times10^{-4}~cm^{3}/mol$ per formula unit that agrees in order of magnitude with
the observed one. \cite{Paschen,Silke} Moreover, the area of the 
Fermi surface is reduced because there are less bands crossing the
Fermi level, what can also be seen in Fig.~\ref{zoom}. As a consequence,
electrical conductivity is expected to be lower and this is also in accord
with the experimental observation. It should be
emphasised that we did not take into account the effect of the
disorder of locked barium atoms. However, due to a very short electron
mean free 
path, which is much smaller than the unit cell parameter, as estimated
from experiments and a simple free electron model, \cite{Silke} the barium
disorder effects should not reflect themselves in
the transport properties. The same argument may be applied to understand
experimental Hall coefficient results, i.e., the electrons might not
be able to close the trajectory on the Fermi surface and therefore
the Hall coefficient might not be strongly influenced by the phase
transition. The absolute value of DOS at the Fermi level is in
agreement with the low temperature specific heat
measurements.\cite{Malte} If we assume the locking of barium atoms at
low temperatures and take the corresponding value of DOS of
25~$eV^{-1}$ at the Fermi level, we obtain $\gamma=$~15~$mJ/molK^{2}$ per
formula unit. The experimental value is 21~$mJ/molK^{2}$ per formula
unit. For the Ba$_{4}$Na$_{2}$Ge$_{25}$, discussed in the next section,
we obtain 29~$mJ/molK^{2}$ per formula unit in comparison with the experimental
33~$mJ/molK^{2}$ per formula unit. The measurements of the variation of
superconducting transition temperature under pressure show that T$_{c}$ is
increased 16 times, from the value of 0.24~K at zero pressure, as the structural phase transition is
suppressed. Introducing the values of DOS for higher and lower symmetry
structure in the McMillan expression for T$_{c}$\cite{McMillan}
provides the ratio of 21, in a relatively good agreement with the
observed one. 

\subsection{The results for Ba$_{4}$Na$_{2}$Ge$_{25}$}

The total band structure for Ba$_{4}$Na$_{2}$Ge$_{25}$ is
qualitatively similar to that found for Ba$_{6}$Ge$_{25}$. The Fermi
level is shifted downwards because there are less electrons and the gap
between bonding and anti-bonding region is narrower since the
lattice parameter is smaller compared to Ba$_{6}$Ge$_{25}$. The peak
of DOS at the Fermi level is also found.

\begin{figure}
\includegraphics[width=.3\textwidth,clip=]{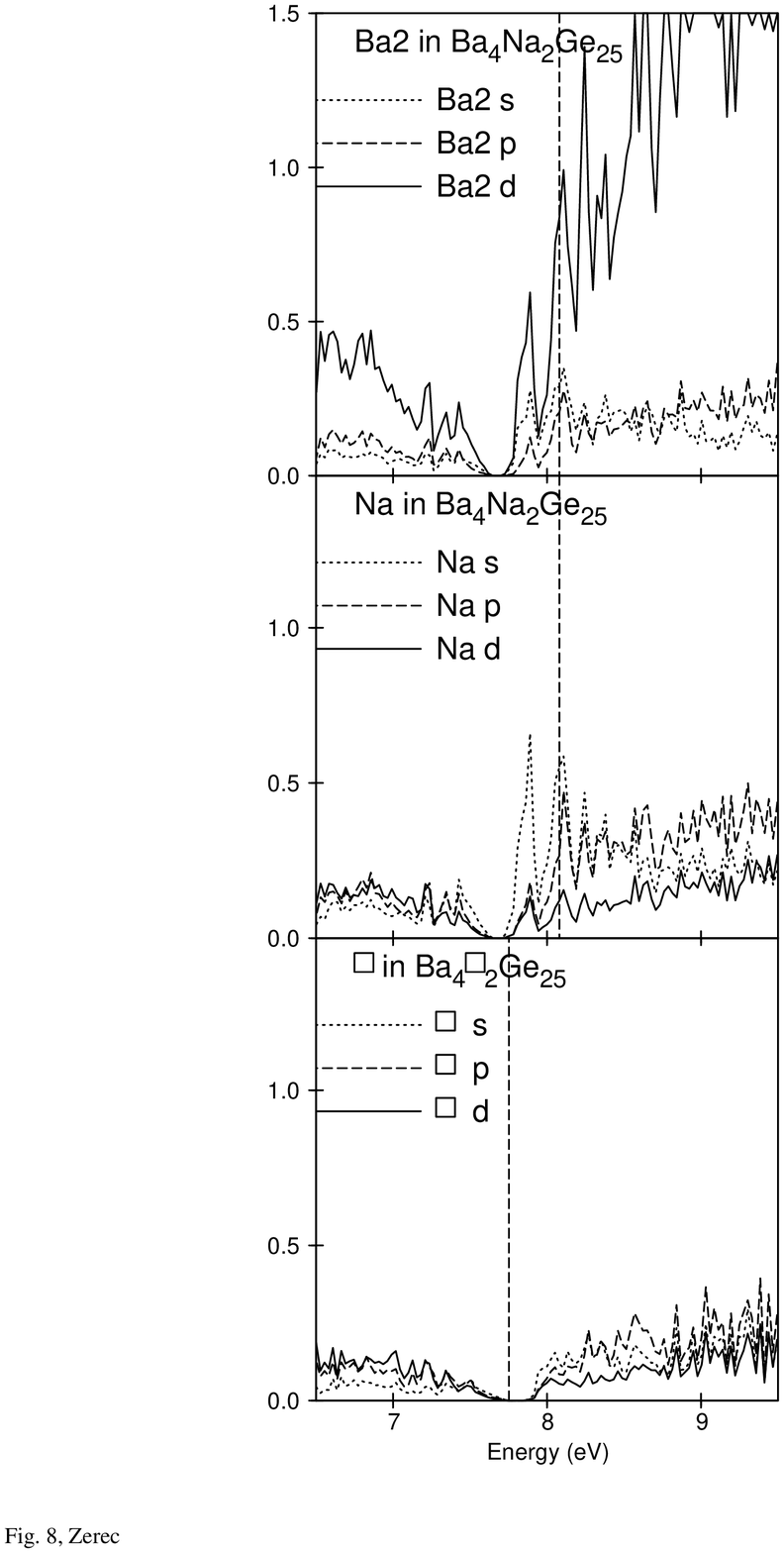}
\caption{\label{ba4e2_dos} The $l$-projected density of states from Ba2
  and Na atomic spheres in Ba$_{4}$Na$_{2}$Ge$_{25}$ and $\Box$
  sphere, equivalent to the Na sphere, in
  Ba$_{4}\Box_{2}$Ge$_{25}$. They all correspond to the Ba2 site in
  Ba$_{6}$Ge$_{25}$ in the original space group No.~213.}
\end{figure}

In order to examine the influence of the sodium on the band structure, we
performed a calculation on a fictitious Ba$_{4}\Box_{2}$Ge$_{25}$
compound and compare it with Ba$_{4}$Na$_{2}$Ge$_{25}$. The inclusion
of sodium atoms introduces a lowering of the bands at the
beginning of the conduction band together with the most upper parts of
the valence bands, so that the fundamental gap has nearly the same
value. The conduction bands are modified. Although some similarities
may be observed, rigid-band model could hardly be applied. In the
valence region, however, apart from the bands just below the
fundamental gap, near $\Gamma$ point, the band structures are almost
identical. Introduction of sodium does not influence the bonding
states of germanium. 
The $l$-projected DOS in sodium atomic spheres are shown in
Fig.~\ref{ba4e2_dos}, and compared with the pertinent empty
sphere. In the sodium sphere a peak of mainly $s$-symmetry at the
Fermi level is found, where the modifications of the band
structure, considered above, is observed.

\begin{figure}
\includegraphics[width=.4\textwidth,clip=]{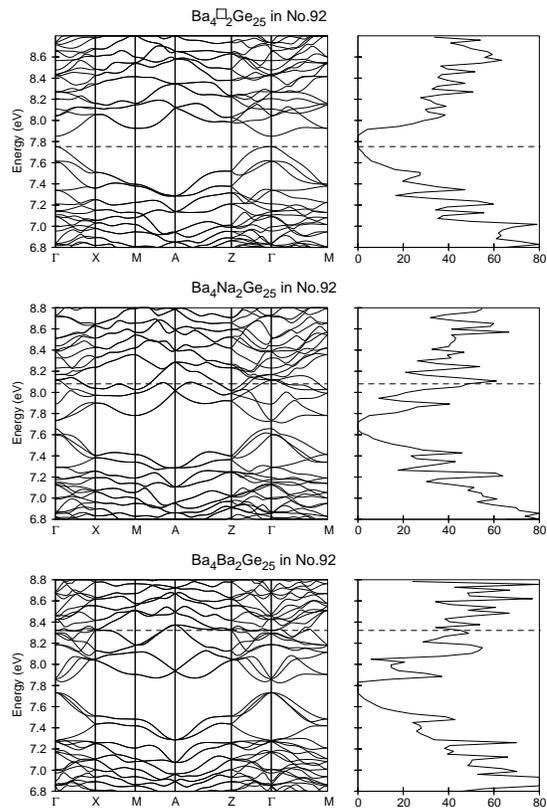}
\caption{\label{ba4e2_com}Comparison of the band structures around the 
  Fermi level of the fictitious Ba$_{4}\Box_{2}$Ge$_{25}$,
  Ba$_{4}$Na$_{2}$Ge$_{25}$, and Ba$_{4}$Ba$_{2}$Ge$_{25}$ with
  same structure and symmetry as Ba$_{4}$Na$_{2}$Ge$_{25}$.}
\end{figure}

We have also performed a calculation on 
Ba$_{4}$Ba$_{2}$Ge$_{25}$, where we have inserted barium instead of
sodium atoms and kept the same symmetry and the structure. It is
interesting that no shifting of the conduction band is
observed. However, the band structure undergoes stronger changes
than in the case of sodium. Even the valence region, i.e.,
bonding germanium states, is significantly affected. 
The $l$-projected DOS in the barium spheres, as in  Ba$_{6}$Ge$_{25}$, 
discovers the appearance of the states of $d$-symmetry
(Fig.~\ref{ba4e2_dos}). The large contribution of $d$-symmetry states
are present around and above the Fermi level, in contrast to the
sodium and the empty spheres, where their contribution is small. 

The electron charges, projected on the states of particular
$l$-symmetry, in the sodium, barium, and empty spheres for
corresponding compounds are given in the Table~\ref{na_charge}. Since
in the empty spheres only the tails from germanium states are present, 
the difference to the spheres of sodium and barium should give the
contribution of more atomic-like states. In the sodium sphere, the
charge is mainly in the states of $s$-symmetry, but comparable
contributions are found in the states of $p$- and somewhat less in 
$d$-symmetry states. Due to the change of the potential in the sodium
sphere in comparison with empty one, it is hard to disentangle the
contribution coming from germanium tails. The values, however, indicate
hybridisation of sodium $s$- and $p$-states with germanium states,
although not so strong as in the case of barium in
Ba$_{4}$Ba$_{2}$Ge$_{25}$, where a distinctively large contribution of
$d$-symmetry states in the barium sphere is found, which again points
to the strong hybridisation of barium $d$-states with germanium $p$-states. 

\begin{table*}
\caption{\label{na_charge} The electron charge in the particular Ba,
  Na, and $\Box$ atomic sphere of Ba$_{4}$Na$_{2}$Ge$_{25}$,
  Ba$_{4}\Box_{2}$Ge$_{25}$, and Ba$_{4}$Ba$_{2}$Ge$_{25}$, decomposed
  in the states of particular orbital angular momentum quantum
  number. The calculations for all compounds were performed within the 
  same space group (No.~92) with the same structure
  parameters. Ba2, Na, and $\Box$ atomic spheres are all of the same
  radius of 4.59~a.u. and are equivalent to the Ba2 site in the
  Ba$_{6}$Ge$_{25}$ in the original space group No.~213.}
\begin{ruledtabular}
\begin{tabular}{l|cccc|cccc|cccc}
    & \multicolumn{4}{c|} {Ba$_{4}$Na$_{2}$Ge$_{25}$} &
\multicolumn{4}{c|} {Ba$_{4}\Box_{2}$Ge$_{25}$} & \multicolumn{4}{c}
{Difference} \\ 
 \raisebox{1.5ex}[-1.5ex]{At.sphere} & $s$ & $p$ & $d$ & $f$ & $s$ & $p$ & $d$ & $f$ & 
 $s$ & p & d & f \\ \hline
Ba2 & 0.53 & 0.76 & 1.52 & 0.43 & 0.49 & 0.76 & 1.51 & 0.43 & 0.04 &
0.00 & 0.01 & 0.00 \\
Na,$\Box$ & 0.65 & 0.89 & 0.71 & 0.39 & 0.34 & 0.65  & 0.58 & 0.37 &
0.31 & 0.24 & 0.13 & 0.02 \\ \hline
\rule[-7pt]{0pt}{19pt} & \multicolumn{4}{c|} {Ba$_{4}$Ba$_{2}$Ge$_{25}$} &
\multicolumn{4}{c|} {Ba$_{4}\Box_{2}$Ge$_{25}$} & \multicolumn{4}{c}
{} \\ \hline
Ba2,$\Box$\footnote{These correspond to the Na site in Ba$_{4}$Na$_{2}$Ge$_{25}$} & 0.55 & 0.76 & 1.53 & 0.43 & 0.34 & 0.65 & 0.58  & 0.37 &
0.21 & 0.11 & 0.95 & 0.06 \\ 
\end{tabular}
\end{ruledtabular}
\end{table*}

A shifting of sodium atoms to one of the split positions did not
produce such a splitting of the band structure as in the case of
Ba$_{6}$Ge$_{25}$, although some reduction of DOS is observed.

\section{Summary and conclusion}

We have performed LDA  band structure calculations for the two
clathrates, Ba$_{6}$Ge$_{25}$ and Ba$_{4}$Na$_{2}$Ge$_{25}$, using
LMTO method within the atomic sphere approximation. The overall band
structure reveals similarities with Ge in diamond type structure and the
comparison with empty Ge cage shows very similar structure in the
bonding region. However, the conduction bands, in
particular at the Fermi level, have a strong contribution of barium
states. The analysis of the $l$-projected DOS shows that these states
are of strong $d$-symmetry character. The calculations yield the peak of
the DOS at the Fermi level that hints on the structural instabilities of the
system. In Ba$_{6}$Ge$_{25}$, shifting of Ba2 atoms, which have the
largest and the most aspherical atomic displacement parameter, toward
one of the experimentally deduced split positions splits the
peak and reduces the DOS. The splitting of the peak at the Fermi
level, due to the shifting of Ba2 atoms and corresponding
symmetry lowering, may therefore be considered as the band Jahn-Teller
effect. Barium in Ba$_{6}$Ge$_{25}$ clathrate can not be considered as
the electron donor only and this compound does not fit into the
simple concept of PGEC. In the case of Ba$_{4}$Na$_{2}$Ge$_{25}$, the
shifting of sodium atoms to one of the split positions does not produce
such a pronounced tendency toward splitting the peak of DOS. If the
locking of Ba2 atoms in Ba$_{6}$Ge$_{25}$ at the phase
transition is assumed, the reduction of DOS at the Fermi level is in
the agreement with experimentally observed abrupt decrease of the
magnetic susceptibility and with the observed variation of superconducting
transition temperature with pressure. It may qualitatively account for
the observed reduction of the electrical conductivity. The specific
heat coefficient $\gamma$ obtained from the low temperature specific
heat measurements is also in agreement with the calculated value of
DOS at the Fermi for Ba$_{6}$Ge$_{25}$ with barium atoms locked,
and for Ba$_{4}$Na$_{2}$Ge$_{25}$ in the average structure.

\begin{acknowledgments}

We wish to acknowledge useful discussions with P. Fulde, S. Paschen,
F. M. Grosche and W. Carrillo-Cabrera.

\end{acknowledgments}


\begin{thebibliography}{32}
\bibitem{Ba6Ge25} W.Carrillo-Cabrera, J. Curda, H. G. von
  Schnering, S. Paschen and Yu. Grin, Z. Kristalogr. NCS \textbf{215},
  207 (2000).
\bibitem{Kim} S.-J. Kim, S. Hu, C. Uher, T. Hogan, B. Huang,
  J. D. Corbett and M. G. Kanatzidis, J. Solid State
  Chem. \textbf{153}, 321 (2000).  
\bibitem{Fukuoka} H. Fukuoka, K. Iwai, S. Yamanaka, H. Abe, K. Yoza
  and L. H\"aming, J. Solid State 
  Chem. \textbf{151}, 117 (2000).
\bibitem{Nolas} G. S. Nolas, G. A. Slack and S.
  B. Schujman, Semiconductors and Semimetals \textbf{69}, 255 (2001).
\bibitem{Paschen} S. Paschen, V. H. Tran, M. Baenitz,
  W. Carrillo-Cabrera, R. Michalak, Yu. Grin and F. Steglich, in
  \textit{Proceedings of XIX International Conference on
    Thermoelectrics}, edited by D. M. Rowe (Babrow Press Wales, UK,
  2000), p. 374. 
\bibitem{Silke} S. Paschen, V. H. Tran, M. Baenitz,
  W. Carrillo-Cabrera, Yu. Grin and F. Steglich, Phys. Rev. B
  \textbf{65}, 134435 (2002).
\bibitem{Malte} F. M. Grosche, H. Q. Yuan, W. Carrillo-Cabrera,
  S. Paschen, C. Langhammer, F. Kromer, G. Sparn, M. Baenitz, Yu. Grin
  and F. Steglich, Phys. Rev. Lett. \textbf{87}, 247003 (2001). 
\bibitem{Na2Ba4Ge25} W. Carrillo-Cabrera, J. Curda, K. Peters
 , S. Paschen, Yu. Grin and H. G. von Schnering,
  Z. Kristalogr. NCS \textbf{216}, 183 (2001).
\bibitem{Grin} S. Paschen, W. Carillo-Cabrera, M. Baentiz,
  V. H. Tran, A. Bentien, H. Borrmann, R. Cardoso Gil,
  R. Michalak, Yu. Grin and F. Steglich, in \textit{MPI-CPfS
    Development of the Institute and Scientific Report}, MPI-CPfS,
  Dresden 2000.
\bibitem{Grin2} W. Carrillo-Cabrera  \textit{et al.}, Phys. Rev. B, to 
  be submitted.
\bibitem{Fassler2} T. S. F\"assler and C. Kronseder,
  Z. Anorg. All. Chem. \textbf{624}, 561 (1998).
\bibitem{Fassler} T. F. F\"assler, Z. Anorg. All. Chem. \textbf{624}, 
  569 (1998).
\bibitem{Muller} U. M\"uller, \textit{Relations between the Wyckoff positions},
in \textit{International tables for X-ray crystalography},
\textbf{A1}, (to be published). 
\bibitem {Andersen} O. K. Andersen, Phys. Rev. B \textbf{12}, 3060
  (1975).
\bibitem{Skriver} Hans L. Skriver, \textit{The LMTO method}, Springer,
  Springer series in solid state sciences 41, Berlin 1984.
\bibitem{Barth} U. von Barth and L. Hedin, J. Phys. C \textbf{5},
  1629 (1972).
\bibitem{Dong} J. Dong and O. F. Sankey, J. Phys.: Condens. Matter, \textbf{11}, 6129 (1999).
\bibitem{Saito} S. Saito and A. Oshiyama, Phys. Rev. B \textbf{51},
  2628 (1995).
\bibitem{Moriguchi} K. Moriguchi, M. Yonemura, A. Shintani and
  S. Yamanaka, Phys. Rev. B \textbf{61}, 9859 (2000).
\bibitem{Moriguchi2} K. Moriguchi, S. Munetoh and A. Shintani,
  Phys. Rev. B \textbf{62}, 7138 (2000).
\bibitem{Blake2} N. P. Blake, D. Bryan, S. Latturner,
  L. M{\o}llnitz, G. D. Stucky and H. Metiu, J. Chem. Phys \textbf{114}, 10063 (2001).
\bibitem{Zhao} J. Zhao, A. Buldum, J. P. Lu and C. Y. Fong,
  Phys. Rev. B \textbf{60}, 14177 (1999).
\bibitem{Nagano} T. Nagano, K. Tsumuraya, H. Eguchi and
  D. J. Singh, Phys. Rev. B \textbf{64}, 155403 (2001).
\bibitem{Demkov} A. A. Demkov, O. F. Sankey, K. E. Schmidt,
  G. B. Adams and M. O'Keeffe, Phys. Rev. B \textbf{50}, 17001
  (1994).
\bibitem{Fukuoka2} H. Fukuoka, K. Ueno and S. Yamanaka,
  J. Organometallic Chem. \textbf{611}, 543 (2000).
\bibitem{Harrison} W. A. Harrison, \textit{Electronic structure
    and the properties of solids}, Dover Publications, Inc., New York
  1989.
\bibitem{Yokoya} T. Yokoya, A. Fukushima, T. Kiss, K. Kobayashi,
  S. Shin, K. Moriguchi, A. Shintani, H. Fukuoka and
  S. Yamanaka, Phys. Rev. B \textbf{64}, 172504 (2001).
\bibitem{Shimizu} F. Shimizu, Y. Maniwa, K. Kume, H. Kawaji,
  S. Yamanaka and M. Ishikawa,
  Phys. Rev. B \textbf{54}, 13242 (1996).
\bibitem{Hagino} T. Hagino, T. Tojo, T. Atake and S. Nagata,
  Phil. Mag. B \textbf{71}, 881 (1995).
\bibitem{Tse} V. I. Smelyansky and J. S. Tse,
  Chem. Phys. Lett. \textbf{264}, 459 (1997).
\bibitem{Brunet} F. Brunet, P. M\'{e}linon, A. San Miguel,
  P. K\'{e}gh\'{e}lian, A. Perez, A. M. Flank, E. Reny,
  C. Cros and M. Pouchard, Phys. Rev. B \textbf{61}, 16550 (2000).
\bibitem{McMillan} W. L. McMillan, Phys. Rev. \textbf{167}, 331 (1968).
\end{thebibliography}
\end{document}